\documentclass[twocolumn,showpacs,showkeys,preprintnumbers,amsmath,amssymb]{revtex4}

\usepackage{graphicx}
\usepackage{dcolumn}
\usepackage{bm}

\begin{document}


\title{Gain/loss asymmetry in time series of individual stock prices and its relationship to the leverage effect}

\author{Johannes Vitalis Siven}
 \email{jvs@saxobank.com}
\author{Jeffrey Todd Lins}%
 \email{jtl@saxobank.com}
\affiliation{%
Saxo Bank A/S, Philip Heymans All\'e 15, DK-2900 Hellerup, Denmark
}

\date{\today}

\begin{abstract}
Previous research has shown that for stock indices, the most
likely time until a return of a particular size has been observed
is longer for gains than for losses. We establish that this
so-called gain/loss asymmetry is present also for individual
stocks and show that the phenomenon is closely linked to the
well-known leverage effect --- in the EGARCH model and a modified
retarded volatility model, the same parameter that governs the
magnitude of the leverage effect also governs the gain/loss
asymmetry.
\end{abstract}

\maketitle

Researchers have estimated empirical distributions for \emph{first
passage times} of financial time series, the smallest time
interval needed for an asset to cross a fixed return level $\rho$.
Jensen, Johansen, and Simonsen \cite{investStatistics} show that
for stock indices, the most likely first passage time is shorter
for $\rho = -5\%$ than for $\rho = 5\%$ --- the first passage time
densities are shifted with respect to each other --- a phenomenon
which they refer to as \emph{gain/loss asymmetry}.

If $\{X_t\}_{t \geq 0}$ denotes the logarithm of a given price
process, for instance daily closing prices of a stock or a stock
index, the first passage time $\tau_\rho$ of the level $\rho$ is
defined as
$$
\tau_\rho = \left\{ \begin{array}{ll}
\min\{s>0;\ X_{t+s} - X_t \geq \rho\}& \mbox{if } \rho> 0,\\
 \min\{s>0;\ X_{t+s} - X_t \leq \rho\}& \mbox{if } \rho< 0,
\end{array} \right.
$$
and is assumed to be independent of $t$. The distribution of
$\tau_\rho$ is estimated in a straightforward manner from a time
series $X_{0},\ldots,X_{T}$. Consider $\rho > 0$, and let $t+s$ be
the smallest time point such that $X_{t+s}-X_t \geq \rho$, if such
a time point exists. In that case, $s$ is viewed as an observation
of $\tau_\rho$. (If $\rho <0$, take instead $t+s$ such that
$X_{t+s}-X_t \leq \rho$.) Running $t$ from $0$ to $T-1$ gives a
set of observations from which the distribution of $\tau_\rho$ is
estimated as the empirical distribution. Given the empirical
distribution, we follow Jensen et al.\ \cite{investStatistics} and
compute a fit of the density function for the generalized gamma
distribution. This density is plotted as a solid line together
with the empirical distribution in all figures, to guide the eye
--- we do not discuss the fitted parameters, nor claim that
$\tau_\rho$ truly follows a generalized gamma distribution.

\begin{figure}
\begin{center}
\includegraphics[width=\columnwidth]{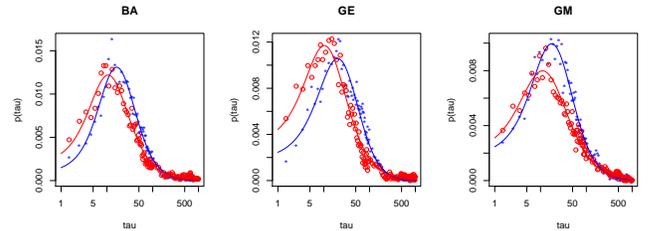}\vspace{-0.2cm}
\end{center}
\caption{Estimated distribution of the first passage time
$\tau_\rho$ for the log price of three individual stocks: Boeing
(BA), General Electric (GE), and General Motors (GM). The graphs
correspond to $\rho = +5\sigma_S $ (stars) and $\rho = -5\sigma_S$
(rings), where $\sigma_S$ is the estimated daily standard
deviation of returns for stock $S$. Values: $\sigma_{BA} = 1.9\%$,
$\sigma_{GE} = 1.7\%$, and $\sigma_{GM} = 2.3\%$.}
\label{fig:GainLoss}
\end{figure}

In an unpublished working paper, Johansen, Jensen, and Simonsen
\cite{inverseMarkets} demonstrate that individual stocks do not
not display a gain/loss asymmetry for $\rho = \pm 5\%$, at
variance with e.g.\ the Dow Jones Industrial Average index. While
we are able to reproduce these results, it is not true in general
that individual stocks do not display gain/loss asymmetry. There
is an asymmetry, but for stocks one has to consider $\rho$ of
greater magnitude than for indices. This is to be expected, since
the standard deviation of daily log returns is typically higher
for individual stocks than for indices. The choice $\rho = \pm
5\%$ corresponds to approximately $\pm 5$ daily standard
deviations for the Dow Jones index --- when $\rho$ is chosen
analogously for the individual stocks they display a clear
gain/loss asymmetry (see Figure \ref{fig:GainLoss}).

\begin{figure}
\begin{center}
\includegraphics[width=\columnwidth]{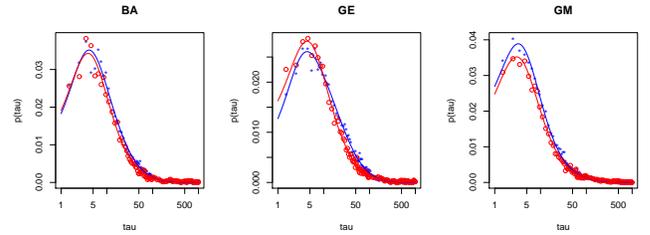}\vspace{-0.2cm}
\end{center}
\caption{Estimated distribution of the first passage time
$\tau_\rho$ for the log price of three individual stocks: Boeing
(BA), General Electric (GE), and General Motors (GM). The graphs
correspond to $\rho = +5\%$ (stars) and $\rho = -5\%$ (rings).}
\label{fig:noGainLoss}
\end{figure}

This finding has some implications for interpreting other results
in the literature. Donangelo, Jensen, Simonsen, and Sneppen
\cite{fearModel} cited the result from Johansen et al.\
\cite{inverseMarkets}, that there is a gain/loss asymmetry for the
Dow Jones stock index but not for the individual stocks, and
proposed the following explanation: sometimes, the stocks move
together, and that this tends to happen for down moves rather than
up moves. Donangelo et al.\ also proposed a probabilistic model,
the \emph{asymmetric synchronous market model}, in which the index
but not the individual stocks display a gain/loss asymmetry.
Siven, Lins, and Lundbek Hansen \cite{multiscale} follow the same
line of thought, but with a more detailed view with regard to the
temporal structure of the phenomenon. They show that the gain/loss
asymmetry in stock indices is a long time scale phenomenon --- it
vanishes if enough low frequency content of the price signal is
removed. They also propose a generalization of the asymmetric
synchronous market model, incorporating prolonged correlations of
the stocks, to account for this fact. While both the asymmetric
synchronous market model and its generalization seem overly
restrictive given that the real stocks in fact \emph{do} display
gain/loss asymmetry, the models prove an important point:
gain/loss asymmetry \emph{can} arise in an index even if it is
absent in the constituents, if the stocks tend to move in a more
correlated manner during downturns. Recent work by Siven and Lins
\cite{temporalStructure} shows that this is indeed the case:
stocks tend to move with a higher degree of dependence in times of
index downturns than in periods of index upturns. That paper also
demonstrates that the gain/loss asymmetry in the Dow Jones index
vanishes if the temporal dependence structure is destroyed by
randomly permuting the returns --- the phenomenon is due to serial
dependence and not properties of the unconditional return
distribution, like skewness. It is straightforward to verify that
this holds true for the individual stocks as well (not reported).


Ahlgren, Jensen, Simonsen, Donangelo, and Sneppen
\cite{frustration} suggest that the gain/loss asymmetry might be
related to the \emph{leverage effect}, the stylized fact that
stocks and stock indices tend to be more volatile in periods
following negative returns. If $\delta X_t = X_t - X_{t-1}$
denotes the stock or index return on day $t$, the leverage effect
can be quantified by $L(\tau) = \textrm{Corr}[\delta X_t,\delta
X_{t+\tau}^2]$, which for stocks and stock indices is found to be
negative and increasing for $\tau \geq 0$, and close to 0 for
$\tau < 0$, see Cont \cite{stylizedCont}. Alghren et al.\ instead
follow Bouchaud, Matacz, and Potter \cite{leverage} and study the
quantity $L^*(\tau) = \textrm{Cov}[\delta X_t,\delta
X_{t+\tau}^2]/\textrm{Var}[\delta X_t]^2$, which is not homogenous
--- scaling the prices (e.g.\ by counting cents instead of
dollars) will change the estimated leverage effect. This is
particularly troubling since Bouchaud et al.\ study the average of
leverage effects for a large number of stocks, all of which surely
do not have the same variance. Below, we use Cont's definition,
$L(\tau) = \textrm{Corr}[\delta X_t, \delta X_{t+\tau}^2]$.
Alternatively, one could consider the quantity $\textrm{Cov}[\delta
X_t,\delta X_{t+\tau}^2]/\textrm{Var}[\delta X_t]^{3/2}$, which is
homogenous, and, additionally, collapses to the unconditional
skewness for $\tau = 0$ --- all our results look qualitatively
similar with this choice.

\begin{figure}
\begin{center}
\includegraphics[width=\columnwidth]{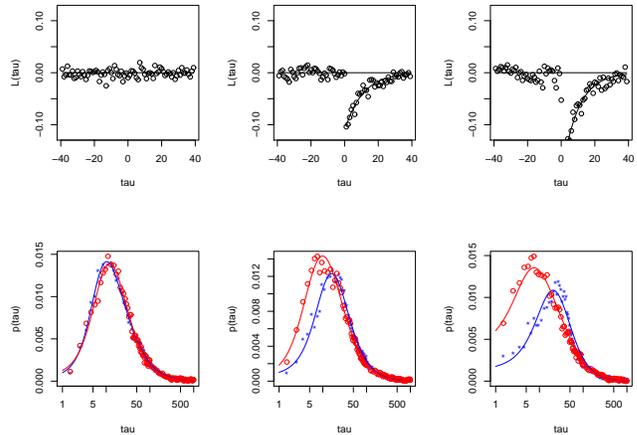}
\end{center}
\caption{EGARCH model with parameters $(\mu,a_0,a_{1b},b_1) =
(0,-0.70,0.20,0.92)$ and $a_{1a} = 0$ (left), $a_{1a} = -0.15$
(middle), and $a_{1a} = -0.30$ (right). Top: Estimated leverage
effect $L(\tau)$. The fitted curves are exponentials
$-Ae^{-\tau/T}$, where $(A,T) = (0.12,12)$ (middle) and $(0.23,8)$
(right). Bottom: Estimated distribution of the first passage time
$\tau_\rho$. The graphs correspond to $\rho = +5\bar{\sigma}$
(stars) and $\rho = -5\bar{\sigma}$ (rings), where
$\bar{\sigma}^2$ is the unconditional variance for each parameter
configuration.} \label{fig:EGARCH1}
\end{figure}

\begin{figure}
\begin{center}
\includegraphics[width=0.67\columnwidth]{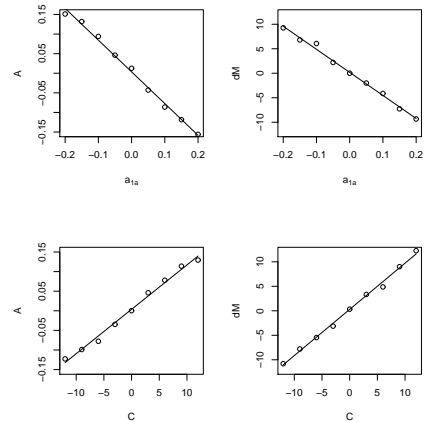}
\end{center}
\caption{Magnitude of leverage effect (measured by $A$ in the fit
$L(\tau) = -Ae^{-\tau/T}$) (left) and the gain/loss asymmetry
(measured by $dM$, the difference between the positions of the
first passage time densities), as function of the parameter
$a_{1a}$ in the EGARCH model (top) and the parameter $C$ in the
modified retarded volatility model (bottom). The remaining
parameters are $(\mu,a_0,a_{1b},b_1) = (0,-0.70,0.20,0.92)$ for
the EGARCH model, and $(\sigma,\alpha) = (0.013,0.90)$ for the
modified retarded volatility model. The gain/loss asymmetry is in
each case estimated for $\rho = \pm 5$ daily standard deviations.}
\label{fig:linear}
\end{figure}

Ahlgren et al.\ \cite{frustration} propose a model of a stock
index, where the individual stocks are driven by a common
stochastic volatility process that supposedly incorporates the
leverage effect. The construction is somewhat forced, no doubt by
the misconception that the gain/loss asymmetry is absent for
individual stocks --- a more serious problem, however, lies in the fact
that their stochastic volatility process is ill-defined and easily
becomes negative. This is unfortunate, since the econometrics
literature contains several thoroughly researched models that
incorporate the leverage effect, see for instance Nelson
\cite{nelson}, Glosten, Jagannathan, and Runkle \cite{GJR},
Zakoian \cite{zakoian}, and Sentana \cite{sentana}. That said,
however, we agree wholeheartedly with Alhgren et al.'s approach to
study the gain/loss asymmetry in models with leverage effect. Our
intuition is in fact that the gain/loss asymmetry and the leverage
effect are closely connected: for $\rho
> 0$, if the log-price process $X_t$ is close to but above the lower
barrier $X_0 - \rho$, then it is likely that we have just
experienced negative returns, so the leverage effect results in
higher volatility and hence higher probability of crossing the
barrier, compared to if $X_t$ is equally close to, but below, the
upper barrier $X_0 + \rho$.

\begin{figure}
\begin{center}
\includegraphics[width=0.67\columnwidth]{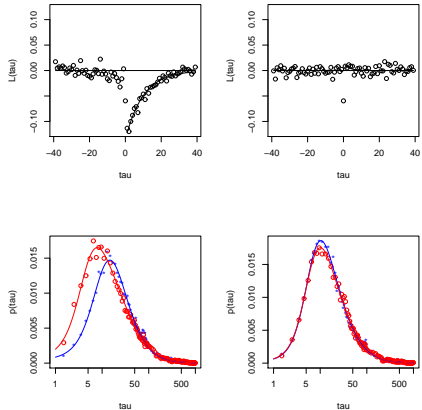}
\end{center}
\caption{Estimated leverage effect (top) and gain/loss asymmetry
(bottom) for a realization of the EGARCH model (left) and a
version of the same realization where the returns have been
randomly permuted (right). The parameters are,
$(\mu,a_0,a_{1a},a_{1b},b_1) = (0,-0.70,-0.15,0.20,0.92)$, and the
bottom graphs correspond to $\rho = +5\bar{\sigma}$ (stars) and
$\rho = -5\bar{\sigma}$ (rings), where $\bar{\sigma}^2$ is the
unconditional variance.} \label{fig:scrambleEGARCH}
\end{figure}

\begin{figure}
\begin{center}
\includegraphics[width=0.67\columnwidth]{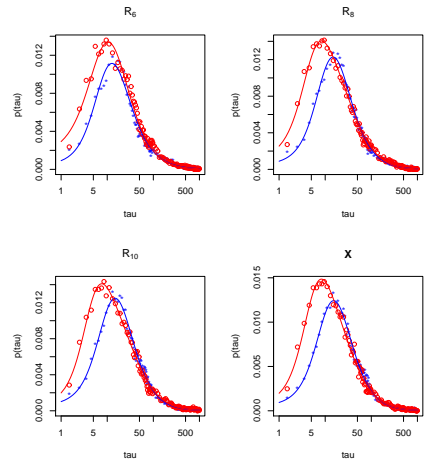}
\end{center}
\caption{Estimated first passage time distributions for a
realization of the EGARCH model with parameters
$(\mu,a_0,a_{1a},a_{1b},b_1) = (0,-0.70,-0.15,0.20,0.92)$, and its
high pass filtrations $R_6,R_8,R_{10}$ --- in $R_k$ the frequency
content corresponding to approximately $2^{k-1}$--$2^{k}$ days and
longer has been removed via a discrete wavelet transform (see
Siven, Lins and Lundbek Hansen \cite{multiscale}). The graphs
correspond to $\rho = +5 \bar{\sigma}$ (stars) and $\rho =
-5\bar{\sigma}$ (rings), where $\bar{\sigma}^2$ is the
unconditional variance, and, incidently, extremely close to the
sample variance of each of the high pass filtrations.}
\label{fig:EGARCHMRA}
\end{figure}

To investigate the strength of this heuristical argument
quantitatively we consider the Exponential GARCH (EGARCH) model
suggested by Nelson \cite{nelson}, which is one of the standard
GARCH-type models that incorporates the leverage effect. In the
EGARCH(1,1) model, the stock return process is defined as $\delta
X_t = \mu - \frac{1}{2}\sigma_t^2 + \varepsilon_t$, where the
logarithm of the conditional variance is specified as $ \log
\sigma_{t}^2 = a_0 + a_{1a}\frac{\varepsilon_{t-1}}{\sigma_{t-1}}
+ a_{1b}\left(\frac{|\varepsilon_{t-1}|}{\sigma_{t-1}} - E\left[
\frac{|\varepsilon_{t-1}|}{\sigma_{t-1}}\right] \right) + b_1\log
\sigma_{t-1}^2$, with $\varepsilon_t \sim N(0,\sigma_t^2)$. Note
that $E[|\varepsilon_{t-1}|/\sigma_{t-1}] = \sqrt{2/\pi}$, and
that $\mu$ is the expected growth rate: the expectation of $S_t =
e^{X_t}$ conditional on $S_{t-1}$ is $E[S_t|S_{t-1}] = e^\mu
S_{t-1}$. The parameter $a_{1a}$ captures the leverage effect. For
``good news'' ($\varepsilon_{t-1}/\sigma_{t-1}
> 0$) the impact of the innovation $\varepsilon_{t-1}$ is
$(a_{1b}+a_{1a})\varepsilon_{t-1}/\sigma_{t-1}$ and for ``bad
news'' ($\varepsilon_{t-1}/\sigma_{t-1} < 0$) it is $(a_{1b} -
a_{1a})\varepsilon_{t-1}/\sigma_{t-1}$. If $a_{1b} = 0$, $\log
\sigma_{t}^2$ responds symmetrically to
$\varepsilon_{t-1}/\sigma_{t-1}$. The unconditional variance
$\bar{\sigma}^2$ of an EGARCH(1,1) can be computed explicitly from
the parameters, see Schmitt \cite[Eqn.\ 6]{EGARCHoption} or
Heynen, Kemna, and Vorst \cite{heynen}.

Figure \ref{fig:EGARCH1} shows the leverage effect and the first
passage time densities corresponding to $\rho = \pm 5\bar{\sigma}$
estimated from realizations of the EGARCH(1,1) model, for $a_{1a}
= 0,-0.15,-0.30$. The drift is set to $\mu = 0$, the other
parameters are taken from Schmitt \cite{EGARCHoption} who estimate
the EGARCH model for German stocks: $(a_0,a_{1b},b_1) =
(-0.70,0.20,0.92)$, and the initial value for the variance process
is set to $\sigma_0^2 = \bar{\sigma}^2$. From the figure we see
that $a_{1a}<0$ gives leverage effect and gain/loss asymmetry
similar to what is typically observed for real stocks and indices,
and that for $a_{1a} = 0$, both phenomena vanish. If $a_{1a}
> 0$ the leverage effect and the gain/loss asymmetry is reversed
--- there is in fact a nice linear relationship between the parameter
$a_{1a}$ and the magnitude of the leverage effect (measured by the
parameter $A$ in the fit $L(\tau) = -Ae^{-\tau/T}$), \emph{and}
between $a_{1a}$ and the gain/loss asymmetry (measured by the
difference between the maxima of the first passage time
distributions) --- see Figure \ref{fig:linear}. We have also
verified that the gain/loss asymmetry and the leverage effect in
the EGARCH model vanish if the returns are randomly permuted
(see Figure \ref{fig:scrambleEGARCH}), and that the gain/loss
asymmetry gradually vanishes if more and more low frequency
content is removed (see Figure \ref{fig:EGARCHMRA}), consistent
with the empirical findings in Siven and Lins
\cite{temporalStructure} and Siven, Lins, and Lundbek Hansen
\cite{multiscale}, respectively.

\begin{figure}
\begin{center}
\includegraphics[width=\columnwidth]{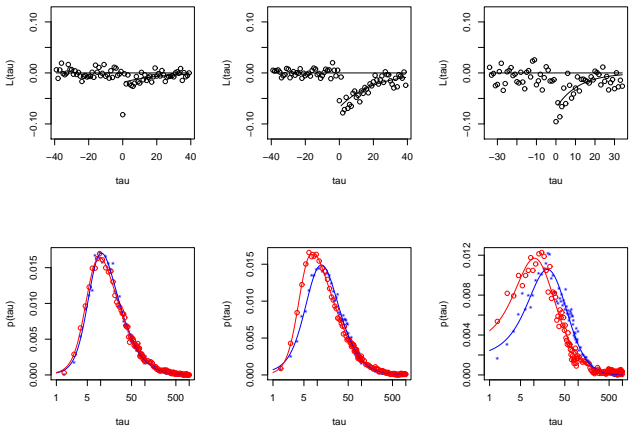}
\end{center}
\caption{Estimated leverage effect (top) and first passage time
densities (bottom) for a realization of Bouchaud et al.'s retarded
volatility model with $(\alpha,\sigma) = (0.985,0.013)$ (left), a
realization of our modified version of that model with
$(\alpha,\sigma) = (0.90,0.013)$ (middle), and the share price of
GE (right). The bottom graphs correspond to $\rho = +5\sigma$
(stars) and $\rho = -5\sigma$, where, for each time series,
$\sigma$ is the daily standard deviation of returns.}
\label{fig:Bouchaud}
\end{figure}

Finally, we consider the \emph{retarded volatility} model by
Bouchaud et al.\ \cite{leverage}, in which the increment $\delta
S_{t} = S_{t} - S_{t-1}$ in the stock price on day $t$ is modelled
as $\delta S_{t} = S^R_t\varepsilon_{t}$, where $\varepsilon_{t}
\sim N(0,\sigma^2)$ and $S^R_t = S_{t-1} - \sum_{\tau = 1}^\infty
\alpha^\tau \Delta S_{t-1-\tau}$, for $\alpha \approx e^{-1/70} =
0.985$. Bouchaud et al.\ study $L^*(\tau) = \textrm{Cov}[\delta
X_{t},\delta X_{t+\tau}^2]/\textrm{Var}[\delta X_t]^2$ and argue
that, approximatively, $L^*(\tau) = -2\alpha^\tau$. Figure
\ref{fig:Bouchaud} shows the estimated leverage effect and first
passage time densities for a realization of the retarded
volatility model: the leverage effect is weak ($A = 0.02$ in the
fit $L(\tau) = -Ae^{-\tau/T}$, compared to $A$ in the order of
0.10 for most stocks), and there is no gain/loss asymmetry.
However, if we follow Qiu, Zheng, Ren and Trimper \cite{qui} and
introduce an additional parameter in the retarded volatility
model, $S^R_t = S_{t-1} - C\sum_{\tau = 1}^\infty \alpha^\tau
\Delta S_{t-1-\tau}$, for $C> 1$ the leverage effect becomes more
pronounced and the model accordingly displays a gain/loss
asymmetry (see Figure \ref{fig:Bouchaud}). Indeed, similarly to
the EGARCH model, there is an approximatively linear relationship
between the parameter $C$, the magnitude of the leverage effect,
and the magnitude of the gain/loss asymmetry (see Figure
\ref{fig:linear}). Note that $C\neq 1$ violates Bouchaud et al.'s
empirical estimate of $L^*(0) = -2$. However, as we remarked
above, we think that this estimate might be compromised by the
lack of homogeniety of $L^*(\tau)$, since they take the average of
leverage effects corresponding to price series with different
variances. On a more fundamental level, we believe that it is
unreasonable for measures of temporal dependence in price
processes to be scale dependent
--- this is obvious at the \emph{very least} for stocks that are traded on
multiple exchanges simultaneously, with prices quoted in the
respective local currencies.

In summary, we have established that individual stocks do indeed
display gain/loss asymmetry, contrary to previous findings, and
observed that randomly permuting the returns of stocks or stock
indices removes the leverage effect as well as the gain/loss
asymmetry. Moreover, in the EGARCH model and a modified retarded
volatility model, the same parameter that governs the leverage
effect also governs the gain/loss asymmetry. These observations
seem to indicate that the gain/loss asymmetry present in stocks as
well as stock indices is an expression of a temporal dependence
structure that is closely related to if not the same as that,
which gives rise to the leverage effect.


\end{document}